\documentclass[]{krakproc}
\usepackage{graphicx}

\def\NoCite#1{}
\def\del#1{}
\def\etal{{\it et al.~}}
\def\eg{{\it e.g.,~}}

\def\lsim{\raise0.3ex\hbox{$<$}\kern-0.75em{\lower0.65ex\hbox{$\sim$}}}
\def\gsim{\raise0.3ex\hbox{$>$}\kern-0.75em{\lower0.65ex\hbox{$\sim$}}}

\newcommand{\crp}{\mathrm{CRp}}
\newcommand{\rmn}{\mathrm}

\def\aj{AJ}
\def\apj{ApJ}
\def\apjl{ApJ Letter}
\def\aap{A\&A}
\def\mnras{MNRAS}

\begin{document}

\title{Magnetic Fields in Clusters of Galaxies}
\author{Torsten A. En{\ss}lin, Corina Vogt, Christoph Pfrommer}
\institute{Max-Planck-Institute for Astrophysics,
Karl-Schwarzschild-Str. 1, 85741 Garching, Germany}
\markboth{T.A. En{\ss}lin}{Magnetic Fields in Clusters of Galaxies}

\maketitle

\begin{abstract}
A brief overview about our knowledge on galaxy cluster magnetic fields
is provided. Emphasize is given to the mutual dependence of our
knowledge on relativistic particles in galaxy clusters and the
magnetic field strength.  Furthermore, we describe efforts to measure
magnetic field strengths, characteristic length-scales, and
power-spectra with reliable accuracy. An interpretation of these
results in terms of non-helical dynamo theory is given. If this
interpretation turns out to be correct, the understanding of cluster
magnetic fields is directly connected to our understanding of
intra-cluster turbulence.
\end{abstract}

\section {What do we know?}

We know that magnetic fields exist in clusters of
galaxies for several reasons. First, in many galaxy clusters we
observe the so-called {\it cluster radio halos} with a spatial
distribution which is very similar to that of the intra-cluster gas
observed in X-rays. These radio halos are produced by
radio-synchrotron emitting relativistic electrons (cosmic ray
electrons = CRe) spiraling in magnetic fields. We do not have direct
evidence of cosmic ray protons (CRp) probably due to their much weaker
radiative interactions. However, in our own Galaxy the CRp energy
density outnumbers the CRe energy density by two orders of magnitude,
which makes the assumption of a CRp population in galaxy clusters very
plausible.

Second, the Faraday rotation of linearly polarized radio emission
traversing the intra-cluster medium (ICM) proves independently the
existence of intracluster magnetic fields. It has been debated, if the
magnetic fields seen by the Faraday effect exist on cluster scales in
the ICM, or in a mixing layer around the radio plasma which emits the
polarized emission (Bicknell \etal 1990, Rudnick \& Blundell
1990)\NoCite{1990ApJ...357..373B, 2003ApJ...588..143R}. However, there
is no valid indication of a source local Faraday effect in the
discussed cases (En{\ss}lin \etal 2003)\NoCite{2003ApJ...597..870E},
and the Faraday rotation signal excess of radio sources behind
clusters compared to a field control sample strongly supports the
existence of strong magnetic fields in the ICM (Clarke et
al. 2001, Johnston-Hollitt \& Ekers 2004)\NoCite{2001ApJ...547L.111C,
2004astro.ph.11045J}.  The detailed mapping of the Faraday effect of
extended radio sources reveals that the ICM magnetic fields are
turbulent, with power on a variety of scales, and with a
power-spectrum which is Kolmogoroff-type (Sect. \ref{sec:RM}). All
these Faraday rotation measurements support magnetic field strengths
of the order of several $\mu$G.

The existence of ICM magnetic fields and cosmic rays is not too
surprising, since there are plenty of energy sources available,
which could have contributed:
{\setlength{\itemsep}{0cm}\setlength{\parsep}{0cm}
\setlength{\topsep}{0cm}\setlength{\parskip}{0cm}
\begin{itemize}\setlength{\itemsep}{0cm}\setlength{\parsep}{0cm}
  \item cluster mergers: shock waves and turbulence (\eg Miniati et
  al. 2000, 2001)\NoCite{2000ApJ...542..608M, 2001ApJ...562..233M},
  \item active galactic nuclei (\eg En{\ss}lin \etal 1997)\NoCite{1997ApJ...477..560E, 1998AA...333L..47E},
  \item injection by galactic winds, driven by supernovae (\eg V{\"o}lk et
  al. 1996)\NoCite{1996SSRv...75..279V},
  \item galactic wakes (\eg Jaffe 1980, Roland 1981, Ruzmaikin et
  al. 1989)\NoCite{1981A&A....93..407R, 1989MNRAS.241....1R,
  1980ApJ...241..925J},
  \item decaying/annihilating dark matter particles
  (\eg Colafrancesco \& Mele 2001, Boehm et
  al. 2004)\NoCite{2001ApJ...562...24C, 2004JPhG...30..279B}.
\end{itemize}
}

\section{Cosmic ray illumination of magnetic fields}

In order to translate cluster radio halo emission into magnetic field estimates one
requires some knowledge about the nature and properties of the
relativistic electron population.  Here, we will discuss the
possibility that the radio emitting CRe are secondaries from hadronic
interactions of a long-living CRp population within the ICM gas (\eg
Dennison 1980, Vestrand 1982, Blasi \& Colafrancesco 1999, Dolag \&
En{\ss}lin 2000, Pfrommer \& En{\ss}lin
2004a)\NoCite{1980ApJ...239L..93D, 1982AJ.....87.1266V,
1999APh....12..169B, 2000A&A...362..151D, 2004A&A...413...17P}:
approximately once in a Hubble time, a CRp collides inelastically with
a nucleon of the ambient ICM gas of non-cooling core clusters. Within
cooling cores, such collisions are much more frequent due to the
higher target densities. Such inelastic proton ($p$) nucleon ($N$)
collisions hadronically produce secondary particles like relativistic
electrons, positrons, neutrinos and $\gamma$-rays according to the
following reaction chain:
\begin{eqnarray}
p + N &\rightarrow& 2N + \pi^{\pm/0} 
\nonumber\\
  \pi^\pm &\rightarrow& \mu^\pm + \nu_{\mu}/\bar{\nu}_{\mu} \rightarrow
  e^\pm + \nu_{e}/\bar{\nu}_{e} + \nu_{\mu} + \bar{\nu}_{\mu}\nonumber\\
  \pi^0 &\rightarrow& 2 \gamma \,.\nonumber
\end{eqnarray}
The resulting $\gamma$-rays can be detected directly with current and
future $\gamma$-ray telescopes. The relativistic electrons and
positrons (summarized as CRes) are visible due to two radiation
processes: inverse Compton scattering of background photon fields
(mainly the CMB, but also starlight photons)
and radio synchrotron emission in ICM magnetic fields.

\section{Hadronic minimum energy criteria}
\label{Pfrommer:MEC}

\begin{figure*}[t]
\begin{tabular}{cc}
\resizebox{0.48\hsize}{!}{\includegraphics{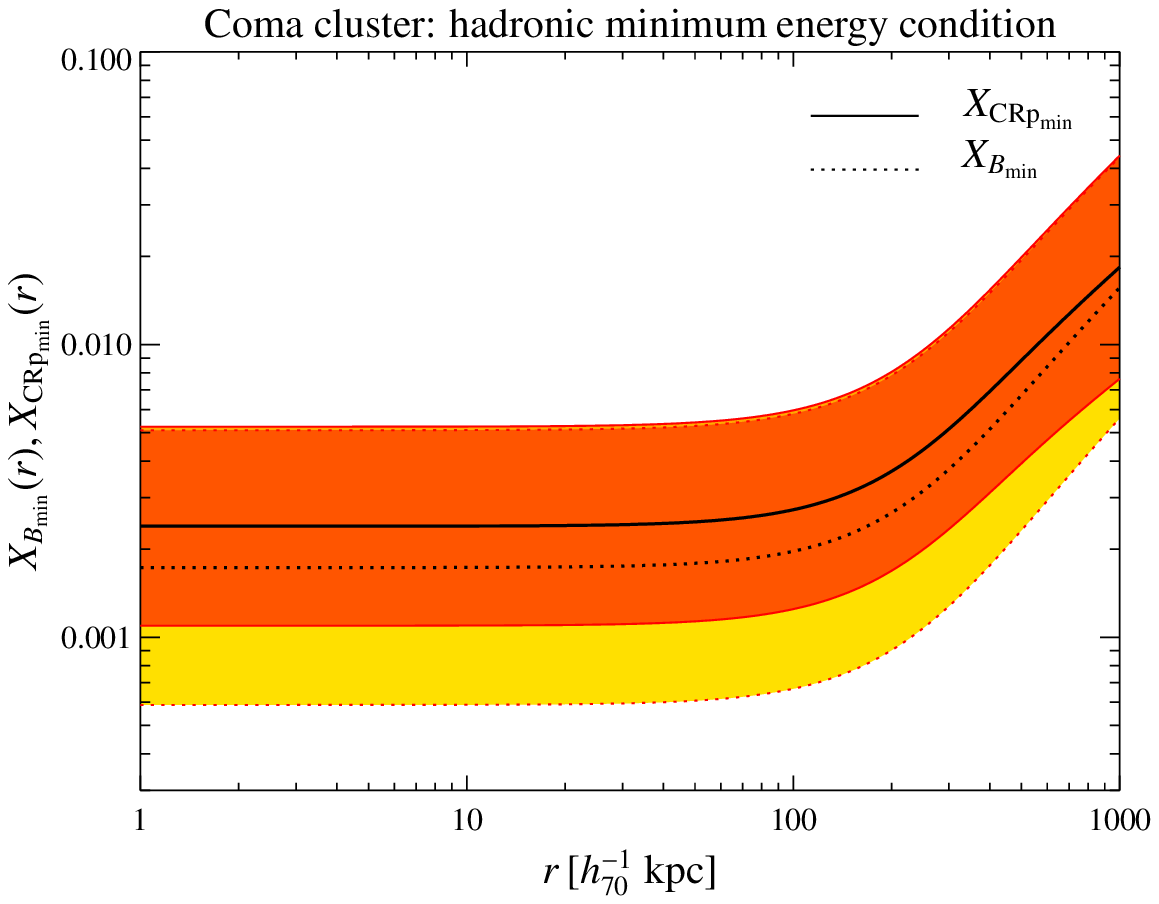}} &
\resizebox{0.48\hsize}{!}{\includegraphics{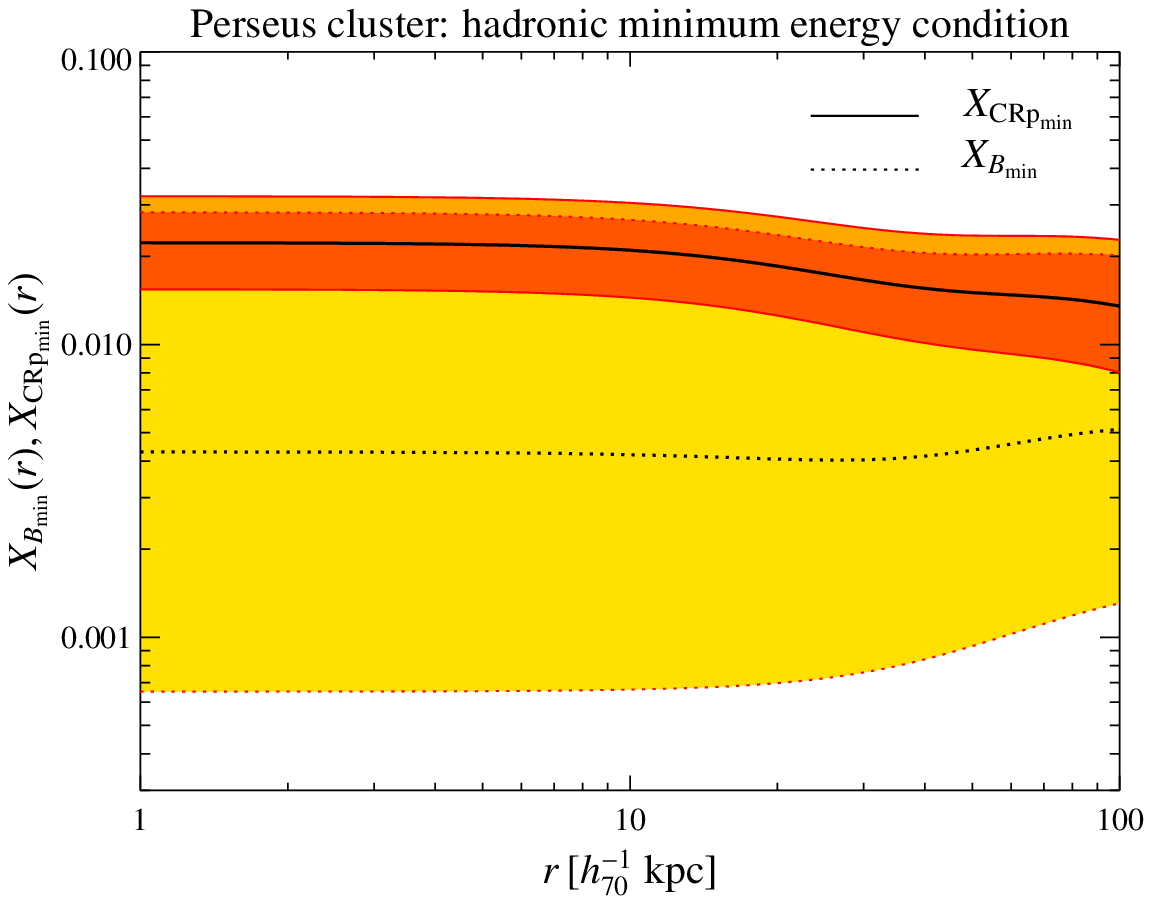}}
\end{tabular}
  \caption{Profiles of the CRp-to-thermal energy density ratio
    $X_{\crp_\rmn{min}}(r)$ (solid) and magnetic-to-thermal energy
    density ratio $X_{B_\rmn{min}}(r)$ (dotted) as a function of
    deprojected radius are shown.  The different energy densities are
    obtained by means of the hadronic minimum energy criterion.  The
    {\bf left panel} shows profiles of the Coma cluster while the {\bf
    right panel} represents profiles of the Perseus cluster.  The
    light shaded areas represent the logarithmic tolerance regions of
    $X_{B_\rmn{min}}(r)$ and $X_{\crp_\rmn{min}}(r)$, respectively,
    while the dark shaded regions indicate the overlap and thus the
    possible equipartition regions in the quasi-optimal case.}
\label{Pfrommer:MECfig}
\end{figure*}

We estimated magnetic field strengths of radio emitting galaxy
clusters by minimizing the non-thermal energy density --- contained in
relativistic electrons, protons, and magnetic fields --- with respect
to the magnetic field strength (Pfrommer \& En{\ss}lin, 2004b).  As
one boundary condition, the implied synchrotron emissivity is required
to match the observed value. Additionally, a second boundary condition
is required mathematically which couples CRps and CRes.  For the {\em
classical} minimum energy criteria, a constant scaling factor between
CRp and CRe energy densities is assumed.  However, if the physical
connection between CRps and CRes is known or assumed, a physically
better motivated criterion can be formulated.  As such a case, we
introduce the minimum energy criterion within the scenario of {\em
hadronically} generated CRes.

Alongside, we provide theoretically expected tolerance regions which measure
the deviation from the minimum energy states by one e-fold: We use logarithmic
measures of the curvature radius at the extremal values in order to
characterize the `sharpness' of the minima.  These regions have the meaning of
a quasi-optimal realization of the particular energy densities.

The philosophy of this approach is to provide a criterion for the
energetically least expensive radio synchrotron emission model
possible for a given physically motivated scenario.  There is no first
principle enforcing this minimum state to be realized in
nature. However, our minimum energy estimates are interesting in two
respects: First, these estimates allow scrutinizing the hadronic model
for extended radio synchrotron emission in clusters of galaxies.  If
it turns out that the required minimum non-thermal energy densities
are too large compared to the thermal energy density, the hadronic
scenario will become implausible to account for the extended diffuse
radio emission. In this respect, our criteria is a way to test the
hadronic model with respect to the observationally available parameter
space spanned by the CRp spectral index and the (unknown) distribution
of the magnetic field strength (Pfrommer \& En{\ss}lin
2004a). Secondly, should the hadronic scenario be confirmed, the
minimum energy estimates allow testing for the realization of the
minimum energy state for a given independent measurement of the
magnetic field strength.

Application to the radio halo of the Coma cluster and the radio mini-halo of
the Perseus cluster yields equipartition between cosmic rays and magnetic
fields within the expected tolerance regions. In the hadronic scenario, the
inferred central magnetic field strength ranges from $2.4~\mu\rmn{G}$ (Coma) to
$8.8~\mu\rmn{G}$ (Perseus), while the optimal CRp energy density is constrained
to $2\% \pm 1\%$ of the thermal energy density (Perseus)
(cf.~Fig.~\ref{Pfrommer:MECfig}). Pfrommer \& En{\ss}lin (2004b) discuss the
possibility of a hadronic origin of the Coma radio halo while current
observations favor such a scenario for the Perseus radio mini-halo.  Combining
future expected detections of radio synchrotron, hard X-ray inverse Compton,
and hadronically induced $\gamma$-ray emission should allow an estimate of
volume averaged cluster magnetic fields and provide information about their
dynamical state.

\section{Faraday rotation and magnetic power spectra}
\label{sec:RM}

\begin {figure*}[ht]
\begin{tabular}{cc}
 \resizebox{0.48\hsize}{!}{\includegraphics{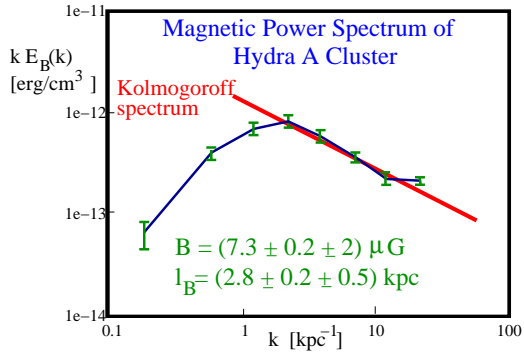}} &
  \parbox[b]{0.48\hsize}{\caption{\label{fig:magpow} Magnetic power spectrum within the
cooling flow region of the Hydra A cluster of galaxies (Vogt \&
En{\ss}lin, submitted). The last data-point at large k-values is
likely affected by map noise and should be discarded.}} 
\end{tabular}
\end{figure*}

The estimates of the magnetic field strength of different methods
differ significantly. Sub-micro-Gauss fields are obtained when
analyzing the reported Extreme Ultraviolet (EUV) and Energy X-ray
(HEX) excesses of the Coma cluster in terms of Inverse Compton (IC)
scattering of CMB photons into theses bands (Lieu \etal 1996,
Hwang 1997, En{\ss}lin \& Biermann 1998, Fusco-Femiano \etal 1996, 2004, but
see Rosetti \& Molendi 2004, who do not find a significant HEX excess
signal). Super-micro-Gauss fields are assumed in the context of the
interpretation of Faraday rotation signals.

Both methods of field estimates have their weak points. The inverse Compton
argumentation can only provide strict lower limits to magnetic fields
strength. The used observed EUV or HEX flux could have (partially) resulted
from a different source or could be a measurement artefact. Therefore, the
number of relativistic electrons could be smaller than assumed in the
estimate, requiring stronger magnetic fields in order to provide the same
amount of observed synchrotron emission.

Faraday rotation based field estimates are also not straightforward, since
magnetic field reversals along the line of sight partially cancel each other's
Faraday signal. What is left as a typical cluster Faraday signal is the result
of a random walk in rotation measure (RM) in the case of turbulent magnetic
fields. The statistical RM signal depends on the statistical magnetic field
strength times the square root of the magnetic autocorrelation length. The
latter is unknown, and thus the Faraday based field estimates suffer from this
uncertainty.  However, the statistical properties of Faraday maps may allow to
measure the magnetic autocorrelation length under relative reasonable
assumptions of statistical isotropy and homogeneity of the magnetic fields
(see En{\ss}lin \& Vogt 2003).

The Faraday rotation based field estimates can
not be accommodated by sub-micro-Gauss field strength. For a typical
cluster like Coma, $3-5\, \mu$G are reproducing the Faraday signal if
a magnetic length-scale of 10 kpc is assumed.  Lowering the magnetic
field strength by one order of magnitude -- as suggested by the IC
based field estimates (in the case of the HEX excess) -- would require
an increase of the magnetic length scale to $\sim$ Mpc in order to
reproduce the Faraday signal strength. But fields ordered on the
cluster size would produce a nearly homogeneous RM signal, and not
exhibit the many sign reversals observed in clusters.


For the Faraday rotation measurements an analysis would be highly
desirable which shows if and how observational artefacts influence
our field estimates. In order to go into this direction, methods to quantify
the level of noise and artefacts in Faraday maps were developed (En{\ss}lin
\etal 2003). They were used to verify the improved quality of Faraday maps and
even the accuracy of Faraday error maps generated with the new {\it
Polarization Angle Correcting rotation Measure Analysis} (PACMAN) algorithm
(Dolag et al. 2004, Vogt \etal 2004)\NoCite{2004astro.ph..1214D,
2004astro.ph..1216V}. These maps were then analyzed with a maximum likelihood
power spectrum estimator (Vogt \& En{\ss}lin, submitted), which is based on
the cross correlation of Faraday signals in pixel pairs, as expected for a
given magnetic power spectrum and galaxy cluster geometry (En{\ss}lin \& Vogt
2003, Vogt \& En{\ss}lin 2003)\NoCite{2003A&A...401..835E,
2003A&A...412..373V}.

The result of this exercise is not only a magnetic power spectrum, which is
corrected for the complicated geometry of the used radio galaxy and of the
Faraday screen, but also an assessment of the errors, and even the cross
correlation of the errors. The power spectrum of the Hydra A cluster cool core
region, which is displayed in Fig. \ref{fig:magpow}, exhibits a
Kolmogoroff-like power law on small scales, a concentration of magnetic power
on a scale of 3 kpc (the magnetic auto-correlation length) and a total field
strength of $7\pm2 \,\mu$G. The given error is the systematic error due to
uncertainties in the Faraday screen geometry. The statistical error is lower
by one order of magnitude.

\section {Is a consistent picture possible?}

Here, we are attempting to draw a consistent picture, which may explain
at least a significant subset of the observational
information:
\begin{itemize}\setlength{\itemsep}{0cm}\setlength{\parsep}{0cm}
  \item the Faraday rotation observations, which point towards turbulent
  fields strength of several $\mu$G strength: in the cool core region of the
  Hydra A cluster a field strength of 7$\mu$G correlated on 3 kpc; in
  non-cooling flow clusters like Coma somewhat lower fields (say 3$\mu$G) with
  a somewhat larger correlation length (say 10-30 kpc). 
  \item the radio halo synchrotron emission of CRe in Coma 
  \item the EUV excess of the Coma cluster, which may be understood as being
  inverse Compton scattered CMB light, and would favor field strength
  about $1.4\,\mu$G (e.g. En{\ss}lin \& Biermann 1998).
\end{itemize}
The Faraday measurements provide us with volume averaged magnetic energy
densities, since the RM dispersion scales as
 $ \langle {\rm RM}^2 \rangle \propto   \langle B^2 \rangle_{\rm
  Vol}$.
The synchrotron emission is also (approximately) proportional to the magnetic
energy density, but weighted with the CRe population around
10 GeV:
 $ L_{\rm radio} \propto   \langle B^2 \, n_{\rm CRe}\rangle_{\rm Vol}$.
Finally, the IC flux is a direct measurement of the number
density of CRe (of the appropriate energy to produce photons of the
observational frequency):
 $ L_{\rm IC} \propto   \langle n_{\rm CRe}\rangle_{\rm Vol}$.
Combining the latter two measurements provides a magnetic field estimate (for
a given or assumed electron spectral slope), which is weighted with the CRe
density:
\begin{math}
\langle B^2 \rangle_{\rm CRe} =   \frac{\langle B^2 \, n_{\rm
    CRe}\rangle_{\rm Vol}}{\langle n_{\rm CRe}\rangle_{\rm
    Vol}} \propto  \frac{L_{\rm radio}}{L_{\rm IC}}  
\end{math}


The magnetic energy density derived from the combination of
synchrotron and inverse Compton flux is significantly lower than the
one derived from RM measurements. This discrepancy might be reconciled
if there is a significant difference between volume and CRe weighted
averages. This would require inhomogeneous or intermittent magnetic
fields, and a process which anti-correlates the CRe density with
respect to the magnetic energy density.  The latter could be
synchrotron cooling in inhomogeneous magnetic fields. In case of an
injection rate of CRe which is un-correlated with the field strength,
as it is expected for the hadronic electron injection, the equilibrium
electron density is
 $ n_{\rm CRe} \propto (B^2 + B^2_{\rm CMB})^{-1}$.
Here $B_{\rm CMB} \approx 3.2 \mu$G describes the field strength
equivalent to the CMB energy density. 

For illustration, we assume that only a small fraction $f_B = 0.1$ of
the volume is significantly magnetized with a field strength of
$10\,\mu$G, and the rest with only $1 \,\mu$G. We will see later that
$f_B = 0.1$ may be a plausible number.  The volume average would give
$\langle B^2 \rangle_{\rm Vol}^{1/2} = 3.3\, \mu$G, whereas the CRe
average gives $\langle B^2 \rangle_{\rm CRe}^{1/2} = 1.5\,
\mu$G. These numbers are in good agreement with the corresponding
field estimates for the Coma cluster based on Faraday rotation and
IC/synchrotron measurements, respectively. A larger ratio in magnetic
field estimates could even be accommodated since the EUV emitting
electrons are at energies below the synchrotron electrons. A spectral
bump of an old accumulated electron population at these energies is
therefore possible, and even plausible due to the minimum in the
electron cooling rate at these energies (Sarazin 1999).


It remains to be shown that there is also a natural mechanism
producing intermittent magnetic fields. The Kolmogoroff-like magnetic
power spectrum in the cool core of the Hydra A cluster indicates that
the magnetic fields are shaped and probably amplified by
hydrodynamical turbulence (\eg De Young
1992)\NoCite{1992ApJ...386..464D}. Therefore, we have to look into the
predictions of the theories of turbulent dynamo theories.

It is generally found by a number of researchers that the non-helical
turbulent dynamo saturates in a state with a characteristic magnetic
field spectrum (\eg Ruzmaikin \etal 1989, Sokolov \etal 1990,
Subramanian 1999 and many others)\NoCite{1989MNRAS.241....1R,
1990IAUS..140..499S, 1999PhRvL..83.2957S}. The effective magnetic
Reynolds number (including magnetic diffusivity due to gas motions caused
by magnetic backreactions) reaches a critical
value of $R_{\rm c} \approx $ 20 ... 60.  The magnetic fields should
exhibit -- more or less pronounced -- the following properties:
\begin{itemize}
\item[A.] The
average magnetic energy density $\varepsilon_B$ is lower than the turbulent kinetic energy
density $\varepsilon_{\rm kin}$ by $\varepsilon_B \approx \varepsilon_{\rm
  kin} \, R_{\rm c}^{-1}$. 
\item[B.] The magnetic fluctuations are concentrated on a scale $l$, which
is smaller than the hydrodynamical turbulence injection scale $L$ by $l
\approx L R_{\rm c}^{-1/2}$.
\item[C.] Correlations exist up to scale $L$, turn there into an
  anti-correlation, and quickly decay on larger scales.
 This may be understood by Zeldovich's flux rope model, in which
  magnetic ropes with diameter $l$ are bent on scales of the order $L$.
\item[D.] Within flux ropes, magnetic fields can be in equipartition with the
  average turbulent kinetic energy density. 
\item[E.] The magnetic drag of such ropes produces a hydrodynamical viscosity
  on large scales, which is of the order of $4\%$ of the turbulent
  diffusivity (Longcope \etal 2003).
\end{itemize}


Turbulent magnetic dynamo theory predicts intermittent magnetic
fields, as favored by the proposed explanation of the discrepancy in
the different magnetic field estimate methods. Let's see if the other
predictions of the theory are in agreement with observations. We
assume, that $R_{\rm c}$ is in the range 20 to 60.
\begin{itemize}
\item[A.] The expected turbulent energy density in the Hydra A cluster
  core is of the order of $(0.3 \ldots 1)\,
  10^{-10}\, {\rm erg\, cm^{-3}}$, which corresponds to turbulent
  velocities of $v_{\rm turb} \approx (300 \ldots 500)$ ${\rm km/s}$. This
  is comparable to velocities of buoyant radio plasma bubbles (En{\ss}lin \&
  Heinz 2002)\NoCite{2002A&A...384L..27E}, which are expected to stir
  up turbulence (\eg Churazov \etal 2001)\NoCite{2001ApJ...554..261C}.
\item[B.] The expected turbulence injection scale in the Hydra A
  cluster core is of the order of $(15 \ldots 25)$ kpc, again consistent with
  the radio plasma of Hydra A being the source of turbulence. The
  dynamical connection of the radio source length scale and the magnetic
  turbulence scale would explain why the Faraday map of Hydra A is
  conveniently sized to show us the peak of the magnetic power spectrum.
\item[C.] Magnetic intermittency in form of flux ropes might have been
  detected as stripy patterns in the RM map of 3C465 (Eilek \& Owen
  2002)\NoCite{2002ApJ...567..202E}.
\item[D.] The fraction of strongly magnetized volume can become as
  small as $f_{\rm B}  = R_{\rm c}^{-1} \approx 0.02 ... 0.05$, a value which is
  more extreme than what we assumed in our example for the Coma cluster.
\item[E.] The expected hydrodynamical viscosity on large scales in the Hydra
  cluster is of the order of $(1\ldots 4) \cdot 10^{28}\,{\rm cm^2/s}$. It is
  interesting to note, that a lower limit on the large scale viscosity of the
  comparable Perseus cluster cool core of $4\cdot 10^{27}\, {\rm cm^2/s}$ was
  estimated by Fabian et al. (2003)\NoCite{2003MNRAS.344L..48F}. An upper
  limit on the viscosity in the (somewhat different) Coma cluster of $\sim
  3\cdot \,10^{29}\, {\rm cm^2/s}$ was derived by Sch{\"u}cker \etal
  (2004)\NoCite{2004A&A...426..387S}. Both limits are consistent with our
  coarse estimate of the large scale viscosity and enclose it.
\end{itemize}

\section{Conclusion}

It should have become clear that the existence of strong and
possibly intermittent magnetic fields (several $\mu$G) in galaxy clusters is
strongly supported by the recent detection of a Kolmogoroff-like magnetic
power spectra in the Hydra A cluster. Some of the discrepancies between
Faraday-based and inverse Compton-based field estimates can be explained by
effects caused by magnetic intermittence, which is expected from turbulent
dynamo theory.

Furthermore, it is argued that the hadronic generation mechanism of
the cluster radio halo emitting electrons is a viable model (among
others). This model is providing a number of stringent predictions
(like minimal gamma ray fluxes, limits on spectral bending, maximal
possible radio luminosities), which allow detailed consistency tests
with future sensitive measurements. If this model is correct, the
concept of hadronic minimum energy estimates can be introduced, and
leads to magnetic field estimates which are well consistent with the
ones derived from Faraday rotation measurements.


\end{document}